\documentclass{aa}
\usepackage{graphicx,txfonts,lscape,longtable}

\newcommand{\HI}{{\ion{H}{i}}}
\newcommand{\HII}{{\ion{H}{ii}}}

\newcommand{\kms}{$\,$km$\,$s$^{-1}$}

\newcommand{\mJybeam}{mJy beam$^{-1}$}
\newcommand{\msun}{{$M_\odot$}}

\newcommand{\sauron}{{\texttt {SAURON}}}

\def\emph#1{{\sl #1}}
\newcommand{\ltsima} {$\; \buildrel < \over \sim \;$}
\newcommand{\gtsima} {$\; \buildrel > \over \sim \;$}
\newcommand{\lta} {\lower.5ex\hbox{\ltsima}}
\newcommand{\gta} {\lower.5ex\hbox{\gtsima}}

\input{psfig}
\begin{document}
\title{ 
Extended, regular \HI\ structures around early-type galaxies}% \\
%relics of their formation\\

\titlerunning{Extended, regular \HI\ structures around early-type galaxies}
\authorrunning{Oosterloo et al.}

\author{Tom A. Oosterloo\inst{1,2}, 
Raffaella Morganti\inst{1,2}, Elaine M. Sadler\inst{3}, \\Thijs van der
Hulst\inst{2} \and Paolo Serra\inst{2}}

\offprints{oosterloo@astron.nl}

\institute{Netherlands Foundation for Research in Astronomy, Postbus
2, 7990 AA, Dwingeloo, The Netherlands 
\and  Kapteyn Astronomical
Institute, University of Groningen, Postbus 800,
9700 AV Groningen, the Netherlands
\and
School of Physics,
University of Sydney, NSW 2006, Australia 
}

\date{Received ...; accepted ...}

\abstract{We discuss the morphology and kinematics  of the  \HI\
 of a sample of 30 southern gas-rich early-type galaxies selected from the
 \HI\ Parkes All-Sky Survey (HIPASS).  This is the largest collection of
 high-resolution \HI\ data of a homogeneously selected sample. Given the
 sensitivity of HIPASS, these galaxies represent the most \HI-rich early-type
 galaxies.  In two-thirds of the galaxies, we find the \HI\ to be in a large,
 regular disk- or ring-like structure that in some cases is strongly
 warped. In  the remaining cases we find the \HI\ distributed in
 irregular tails or clouds offset from the galaxy.  The giant, regular \HI\
 structures can be up to $\sim 200$ kpc in diameter and contain up to $10^{10}
 M_\odot$ of \HI. The incidence of irregular \HI\ structures appears to be
 somewhat higher in elliptical galaxies, but the large, regular structures are
 observed in both elliptical and S0 galaxies and are not strictly connected to
 the presence of a stellar disk. If these two types of galaxies are the result
 of different formation paths, this is not strongly reflected in the
 characteristics of the \HI.\\ The size and the regular kinematics of the \HI\
 structures imply that the neutral hydrogen must have settled in these
 galaxies several Gyr ago.  Merging as well as gas accretion
 from the IGM are viable explanations for the origin of the gas in these
 galaxies. The average column density of the \HI\ is low so that little star
 formation is expected to occur and these early-type galaxies can remain gas
 rich for very long periods of time. The large \HI\ structures likely
 represent key structures for tracing the origin and evolution of these
 galaxies.

\keywords{galaxies: elliptical and lenticular - galaxies: ISM - galaxies:
neutral hydrogen - radio lines: galaxies}   
}
\maketitle

\section{Introduction}

Recent studies of nearby field elliptical and lenticular galaxies have
revealed a variety of characteristics that can be considered to be signatures
of the continuing assembly of this class of objects. For example, optical
studies show that most early-type galaxies contain significant amounts of
ionised gas and that in most cases this gas must have been accreted (e.g.\
Sadler \& Gerhard 1985, Sarzi et al.\ 2006 and references therein). Other
evidence for continuing accretion and merging is the common occurrence of fine
structure in the optical morphology (Malin \& Carter 1983, Schweizer \&
Seitzer 1992, Schweizer 1998). In addition, detailed stellar-population
studies show that many systems do contain a (often small) subpopulation of
relatively young stars that may have formed from accreted material (see e.g.\
Trager et al.\ 2000, Tadhunter et al.\ 2005). Taken together, the above
results strongly suggest that many field early-type galaxies continue to grow
and that the accretion of gas plays an important role in this.

The relevance of gas in the formation and evolution of early-type galaxies is
also suggested by theoretical work that indicates that dissipative mergers and
accretions are needed to explain the dynamical structure of, in particular,
the more disky early-type galaxies (e.g.\ Naab, Kochfar \& Burkert
2006). Having a good inventory of the gas properties of early-type galaxies is
therefore essential for understanding the formation and evolution of
these systems.

Already many years ago, it was realised that gas can be important for the
evolution of early-type galaxies. Based on compilations of published (mainly
single-dish) data, e.g.\ Knapp, Turner \& Cunniffe (1985) and Roberts \&
Haynes (1994) showed that a significant fraction (5-10\%) of E/S0 galaxies
contain a considerable amount of \HI.  They also showed that, in contrast to
spirals, the \HI\ content of early-type galaxies is not correlated with their
optical luminosity.  This is often taken as evidence for the origin of the gas
being external. However, the situation must be more complex because it is
clear that also many spiral galaxies (including our own Galaxy and M31) are
still accreting small companions and gas (van der Hulst \& Sancisi 2004;
Westmeier, Braun \& Thilker 2005) so that the lack of correlation between the
gas content and the optical luminosity in early-type galaxies must also
involve other factors, such as where the accreted gas ends up in the galaxy,
or the way in which the accreted gas is consumed (or not) by star formation.

In early-type galaxies that are gas rich, the \HI\ is expected to have low
column density and to be distributed over a large region because if it were
not, the column densities would be high enough for large-scale star formation
to occur and the galaxy would not be classified as an early-type. Hence, \HI\
imaging studies are likely to give information on gas properties on scales
larger than the optical galaxy and are therefore a useful complement to
optical work that in most cases investigates the regions inside the optical
boundaries of galaxies. It is at large radii where the dynamical timescales
are larger and where, in particular for field galaxies, the environment is
relatively quiet. Hence, the signatures of the formation history of a galaxy
can survive there for longer periods of time. In this respect, observations of
large-scale gas structures can be important because they may give information
over long timescales.

The most commonly considered scenarios for the formation of early-type
galaxies do indeed predict that such large-scale gas systems may form in some
early-type galaxies for certain initial conditions.  Numerical simulations of
gas-rich mergers show, in particular when the angular momenta of the galaxies
and of the orbit are more or less aligned, that some fraction of the gas of
the progenitor galaxies is at first distributed along very extended tidal
structures that at a later stage can fall back to settle into a large disk
around the newly formed galaxy (e.g.\ Barnes 2002, Naab et
al.\ 2006). Alternatively, galaxies can accrete gas from the IGM; simulations
show that although most of this gas is heated by shocks to the virial
temperature of the halo, some fraction stays below $10^5$ K and can accrete
onto a galaxy as cold gas and may form gaseous structures (e.g., Keres et al.\
2005; Macci\`{o}, Moore \& Stadel 2005). If the galaxy density of environment
is relatively low, the structures of accreted gas will not be destroyed by the
passage of neighbouring galaxies, and both merging and cold accretion can lead
to the formation of large stable and massive structures of cold gas around
early-type galaxies.

Here we present an analysis of high-resolution \HI\ data of a large,
homogeneously selected  sample of early-type galaxies based on  the \HI\
Parkes All Sky Survey (HIPASS, Barnes et al.\ 2001).  HIPASS is a
\HI\ single-dish survey of the entire southern sky with an RMS noise level  of
13 \mJybeam\ over 18 \kms\ and is well suited for constructing homogeneously
selected samples that can be the basis for statistical studies of global
properties.  In Paper II (Sadler et al.\ 2006) we report on a statistical
study focused on the global \HI\ properties of early-type galaxies.  It is,
however, crucial to also investigate the morphology and kinematics of the \HI\
gas in early-type galaxies with higher spatial resolution. Only through the
study of the detailed characteristics (morphology, column density, and
kinematics) one can hope to understand the origin of this gas and the role it
may have played in the evolution of the galaxy, while one can also investigate
the effects of confusion in the HIPASS data.  Therefore, we carried out ATCA
follow-up observations of the HIPASS detections and these are the subject of
this paper. Despite its potential importance, such detailed information is
available only for a limited and inhomogeneous sample of early-type galaxies
(e.g.\ Schiminovich et al.\ 1994, 1995; van Gorkom \& Schiminovich 1997;
Morganti et al.\ 1997, Balcells et al.\ 2001; Oosterloo et al.\ 2002), and
only recently systematic studies have begun to shed light on this relatively
unexplored field. Morganti et al.\ (2006) carried out a small but deep \HI\
survey of nearby E/S0's in the northern hemisphere. They observed a subset of
the \sauron\ sample of early-type galaxies (de Zeeuw et al.\ 2002) with the
Westerbork Synthesis Radio Telescope (WSRT). These observations are much
deeper than most earlier work with detection limits of about few $\times
10^6$\msun. Most interestingly, they found that a very high fraction (70\%) of
these systems contains \HI, with morphologies ranging from small clouds to
large, regular disks of low column density \HI. This detection rate is very
similar to that of the ionised gas in the \sauron\ sample (Sarzi et al.\
2006).  These high detection rates underline the importance of gas in the
evolution of early-type galaxies.

The paper is structured as follows: we explain the selection of the sample in
Sec.\ref{sample}; we describe the observations in Sec.\ref{observations}; we
show the result of the observations in Sec.\ref{results}; we discuss these
results and compare them to those obtained by  Morganti et al.\ (2006) in
Sec.\ref{discussion}; finally, we summarise the main conclusions in
Sec.\ref{conclusions}.

\section{Background of this study and selection of the sample}
\label{sample}

The selection of the sample of gas-rich early-type galaxies was done in 2000
and predates the construction of the final HIPASS catalogue (Meyer et al.\
2004). Instead, the selection was done on the HIPASS data cubes themselves.
As also described by Sadler et al.\ (2006), we used the HIPASS data cubes to
create co-added \HI\ spectra at the position of each E and S0 galaxy that is
listed in the Third Reference Catalogue of Bright Galaxies (RC3; de
Vaucouleurs et al.\ 1991) and is south of declination $-30^\circ$. We chose
the RC3 because it contains reliable galaxy Hubble-type classifications based
on uniform, high-quality plate material. This minimises the risk of
contaminating our E/S0 galaxy sample with spirals. The co-added HIPASS spectra
were inspected by two of us (TAO and EMS).  The main criteria for identifying
a candidate \HI\ detection were as follows:
\begin{itemize} 
\item 
An integrated \HI\ flux of at least 2.6\,Jy\,\kms\ in the HIPASS spectrum at 
the position of the RC3 galaxy. 
\item 
A measured \HI\ velocity within 200 \kms\ of the catalogued optical velocity. 
\end{itemize}

The 2.6\,Jy\,\kms\ cutoff is fairly high and will exclude some genuine weak
detections, but was chosen because the spectral baselines of the HIPASS data
are, in some cases, not flat due to difficulties in calibration.  A lower
detection limit would cause contamination by noise peaks and would make the
statistics less reliable.  The requirement for a match between \HI\ and
optical velocities was similarly chosen to exclude spurious \HI\ detections.
Over 90\% of the RC3 galaxies in our sample had a measured optical velocity at
the time the HIPASS selection was done.

The HIPASS detection rate for galaxies with systemic velocities above 6500
\kms\ was significantly lower than for the rest of the sample, while galaxies with
velocities less than 500 \kms\ can be lost in Galactic or HVC \HI\
emission. The final sample is therefore restricted to objects with $500 <
V_{\rm hel} < 6500$ \kms. 

There are 132 E and 533 S0 galaxies in the RC3 which have optical velocities
in this range and the cross-correlation with HIPASS resulted in many candidate
\HI\ detections. Thirteen E and 73 S0 galaxies were marked as such.  Of these
galaxies, for 3 E and 9 S0 galaxies good data are available in the literature
and were therefore not re-observed (see Table 4). All 10 remaining E galaxies
were observed in our ATCA follow up. Of the 64 remaining S0 galaxies, we
observed with ATCA only those 32 for which optical images showed that more
than one galaxy is present in the Parkes beam and hence that the HIPASS data may
suffer from confusion with companion galaxies. If there is some correlation
between the \HI\ properties of S0 galaxies and their environment, this
selection could imply some selection effect. 

The final list of 42 objects observed with ATCA, grouped in E and S0, is given
in Table~1. Information about the 12 galaxies of our sample for which the \HI\
properties were already known from the literature is given in Table 4.

\section{The ATCA observations}
\label{observations}

In total 42 objects were observed with ATCA between May 2001 and January
2002. All observations were performed using the 375-m array configuration and
256 channels covering a 16-MHz wide band centred on the frequency of the
redshifted hydrogen line. In order to limit the total amount of observing
time, several galaxies were observed within a single 12-hr observing slot.
For most galaxies the data consist of several short cuts spread over 12 hrs to
provide, given the relatively short integration time, a sufficient, albeit not
optimal, $uv$-coverage for imaging the field. This resulted in typical
integration times of 2 hours per galaxy and this is the main limitation to the
quality of the images. For a small number of galaxies the scheduling allowed
longer integration times and better $uv$ coverage (see Table 1 for
details). In a few datasets (in particular ESO~92--21 and NGC~6920) some \HI\
emission is likely not detected because of RFI (Radio Frequency Interference)
around 1408 MHz.

The calibration of the spectral data was done with the MIRIAD package (Sault
et al.\ 1995), which has several features particularly suited for ATCA data,
and was done following the standard recipes for ATCA data.  The flux density
scale was set by observations of PKS 1934--638, for which we adopted the
standard spectrum as given in the ATCA manuals (which is, to zeroth order, a
flux density of 14.9 Jy at 1400 MHz).  This source was also used as bandpass
calibrator.  Spectral-line data sets were constructed by subtracting the
continuum emission using a linear fit through the line-free channels of each
visibility record and subtracting this fit from all the frequency channels
(i.e.\ using the task UVLIN of MIRIAD). This also resulted in continuum
datasets that were processed separately (see below).  The final cubes were
made with natural weighting.  The velocity resolution of the data cubes is,
after Hanning smoothing, about 26 \kms. A summary of the integration times
and noise levels is given in Table~1. The typical FWHM of the restoring beam
is $130\times100$ arcsec$^2$ and the typical noise level is  2--3
\mJybeam.

The moment analysis of the data cubes was also done in MIRIAD.  The total
intensity images of the \HI\ emission were derived based on masks determined
from data cubes made by spatially smoothing the original cube to a resolution
about twice lower than the original.  Pixels with signal below $3\sigma$ in
the smoothed cube were set to zero in the original cube (van Gorkom \& Ekers
1989). Figs 1 and 2 show the total \HI\ contours of the detected galaxies
superimposed on an optical image.  For galaxies were no \HI\ was detected,
the upper limits on \HI\ emission were calculated as three times the
statistical error on having no signal over a width of 200 \kms\ and over one
synthesised beam.  The \HI\ masses (or upper limits to them) are given in
Table~2 together with the velocity width of the integrated \HI\
profile\footnote{We adopt through this paper an Hubble constant of $H_{\circ}
= 70$ km s$^{-1}$ Mpc$^{-1}$}.  In the end, of the 54 galaxies of our sample,
30 turn out to have \HI\ associated with them.

The continuum data were also reduced using MIRIAD. The final continuum images
were made with uniform weighting. The RMS noise levels typically range between
1 and 2 \mJybeam\ (although there are a few cases with much higher noise).
All the detected sources appear unresolved and their fluxes (as well as the
upper limits for the non-detections) are given in Table~2. For some sources,
the poor $uv$-coverage resulted in relatively low image quality and this
increases the uncertainty of the fluxes given in Table 2.

In almost every field observed, objects other than the target galaxy were
detected in \HI. These objects are often close to the target and therefore are
likely companion galaxies. This information is particularly useful for the
study of the effects of the environment that is believed to be an important
factor in the evolution of early-type galaxies. The results from the analysis
of the environment will be discussed in a forthcoming paper.

\begin{figure*}
\label{disks}
\centerline{\psfig{figure=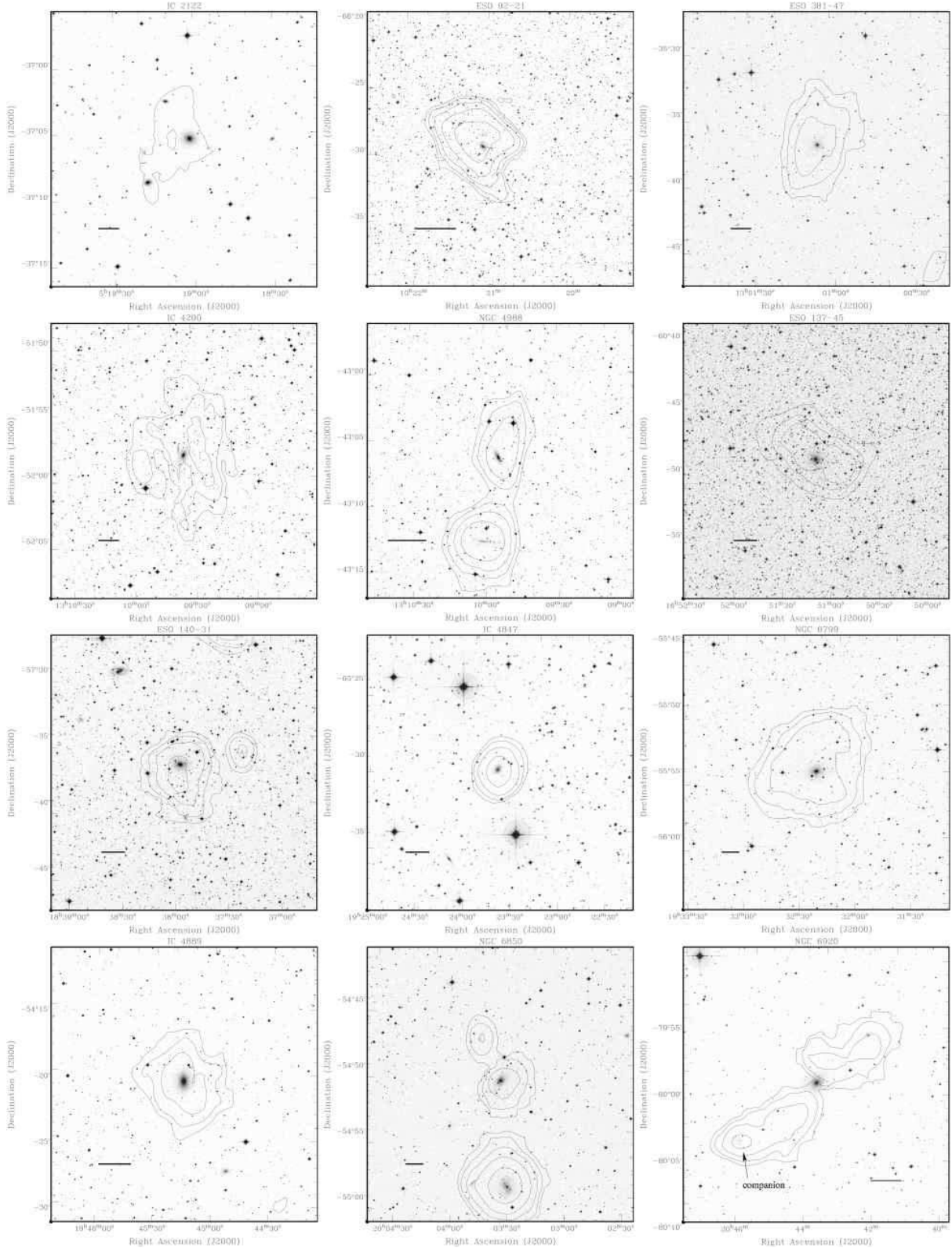,width=16cm}}

\caption{Total intensity images of the fields where \HI\ has been
            detected distributed in a disk-like structure around the target
            galaxy.  The size of fields is $20\times20$ arcmin for all
            galaxies. The horizontal bar indicates 25 kpc.  The target galaxy
            is in the field centre. The contour levels are given in the last
            column of Table 2. In some datasets (in particular ESO~92--21 and
            NGC~6920), some channels likely to contain \HI\ emission were
            affected by RFI therefore in these cases we may be missing some
            emission. For ESO92--21 the \HI\ likely extends more to the SW
            while in NGC 6920 there is likely more \HI\ near the galaxy.}
\end{figure*}

\begin{figure*}
\label{tailsetal}
\centerline{\psfig{figure=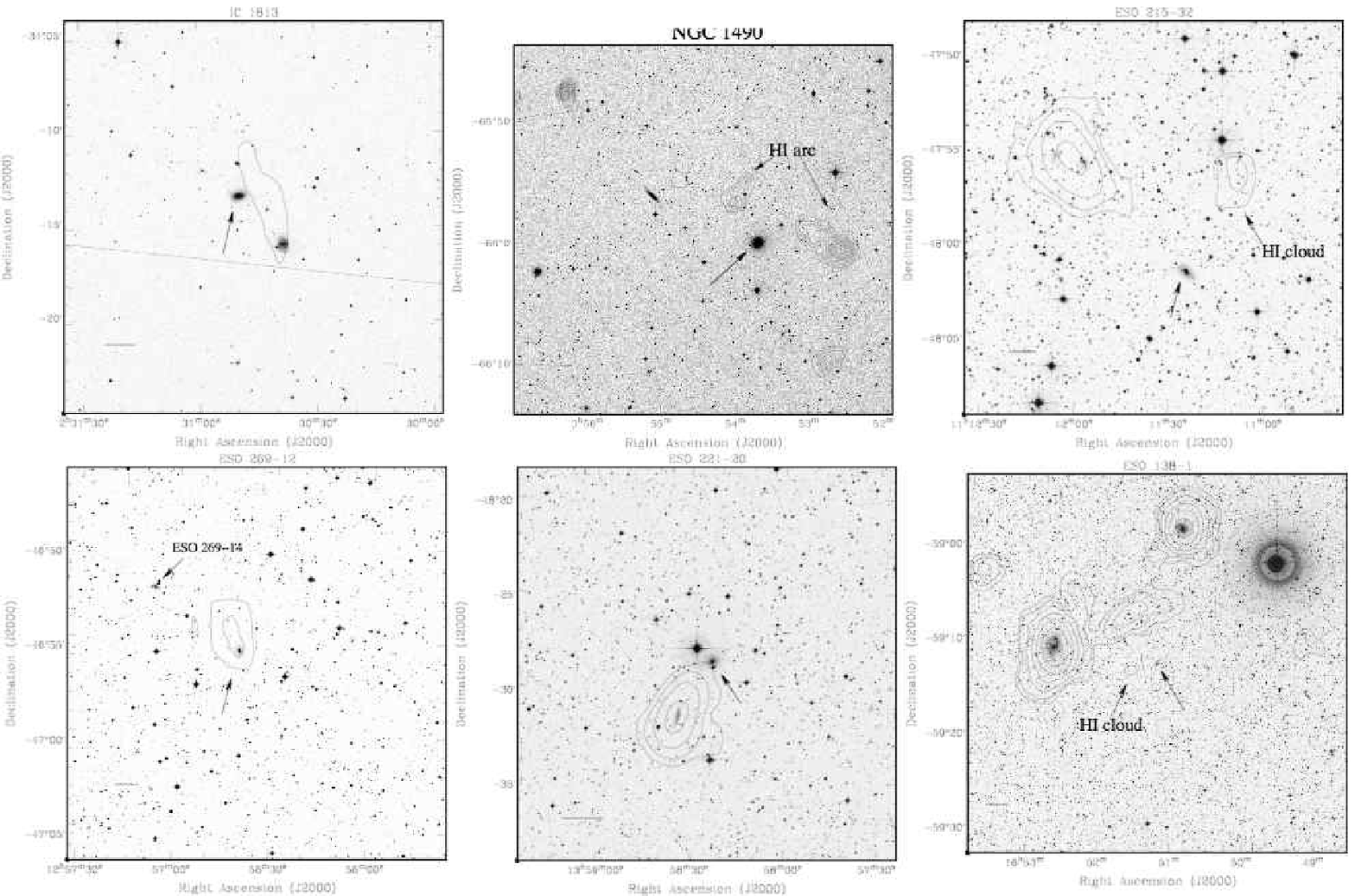,width=16cm}}

            \caption{Total intensity images of the fields where \HI\ has been
            detected in a tail-like structure or offset from the target
            galaxy.  The size of fields is $20\times20$ arcmin, except for NGC
            1490 and ESO 138-1 where a larger area is plotted.  The
            horizontal bar indicates 25 kpc. The target galaxy is marked. The
            contour levels are given in the last column of Table 2. The small
            \HI\ cloud related to ESO 138--1 is indicated by the arrow. The
            other, much brighter, \HI\ in this figure is related to 2
            foreground spirals that are unrelated to ESO 138--1.}
            \end{figure*}

\section{Results}
\label{results}

\subsection{\HI\ detections and their morphologies}

Despite the limitations of the current data (sparse $uv$ sampling and limited
sensitivity), several interesting results are found. In 30 galaxies the
kinematics and/or morphology suggests that the \HI\ has some relation to the
galaxy (18 in our new ATCA data and 12 cases from the literature). This
includes objects where \HI\ is detected  at the position of
the target galaxy {\sl or} it is offset from the galaxy but not obviously associated to any
other galaxy in the field and it has a velocity close to that  of the
target.

\begin{figure*}
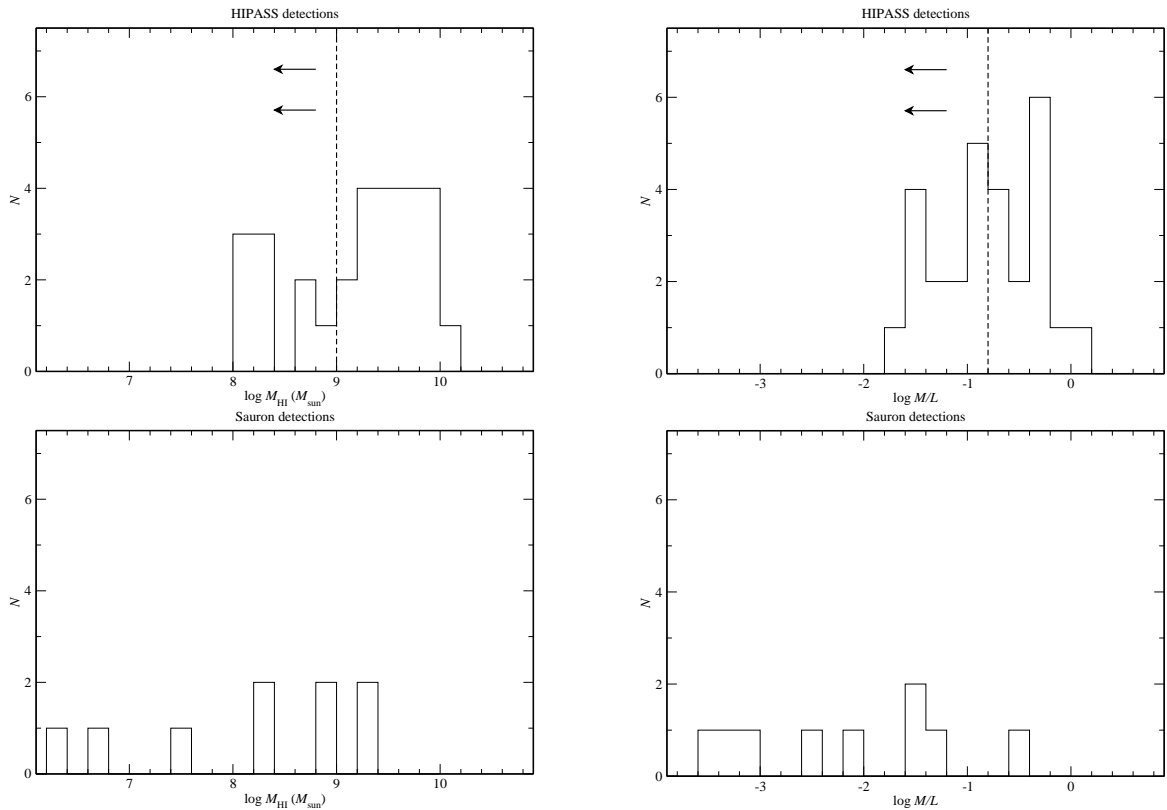

\label{histo}
\centerline{\hfill\psfig{figure=massesAllDetHist.eps,angle=0,width=7cm}
           \hfill \psfig{figure=mlAllDetHist.eps,angle=0,width=7cm}\hfill}
\centerline{\hfill\psfig{figure=massesSauDetHist.eps,angle=0,width=7cm}
            \hfill\psfig{figure=mlSauronDetHist.eps,angle=0,width=7cm}\hfill}
\caption{Distribution of \HI\ mass and $M_{\rm HI}/L_B$ 
for the \HI\ detections (both our observations and galaxies from
literature).The approximate mass detection limit, based on our selection
limit and a characteristic redshift of 3000 km s$^{-1}$ is indicated. The
vertical line in the $M_{\rm HI}/L_B$ histogram indicates the region where all
upper limits to $M_{\rm HI}/L_B$ lie.  For comparison, the distributions of
the same parameters from the much deeper observations of the \sauron\ sample
of galaxies (Morganti et al.\ 2006) are also shown}
\end{figure*}

Due to the limited sensitivity of the HIPASS catalogue, the objects discussed
here are extremely \HI-rich objects. Our selection limit of 2.6 Jy km s$^{-1}$
corresponds to a \HI\ mass of about $10^9$ $M_\odot$ for a redshift of 3000 km
s$^{-s}$. Many galaxies in our sample have a higher redshift so our sample is
incomplete for smaller masses. Therefore, the class of objects
discussed here is somewhat different than those detected in the study of the
\sauron\ galaxies (Morganti et al.\ 2006) because that is based on much more
sensitive observations. This is also evident from the histograms in Fig.\ 3
showing the distribution of the \HI\ masses of the detections of the two
samples.  The \HI\ masses of the galaxies detected in our study range from a
few times $ 10^8$ to more than $10^{10}$ $M_\odot$ (i.e., several times that
of the Milky Way).

A look at the \HI\ morphology reveals that in most of the detections (20
out of 30) the \HI\ is distributed in a low column density disk or ring-like
structure centred on the optical galaxy. Most of these large structures have
regular kinematics. This is illustrated by the position-velocity diagrams
taken along the kinematical major axis that are shown in Fig.\ 4. The
relatively high incidence of these large structures and their characteristics
represents the most important result of these observations. A smaller group
of detections (e.g.\ ESO~221--20 and IC~1813) show tail-like structures
pointing away from the galaxy or to the galaxy from a nearby companion. For
ESO~269--12 it is uncertain whether it contains a disk or whether it has a
tail toward the companion ESO~269--14.  Another small subset of galaxies
(e.g.\ NGC~1490, ESO~138--1 and ESO~215--32) shows cloud-like \HI\ structures
detected (many tens of kpc) away from the target with no obvious optical
counterpart but with velocities similar to the systemic of the target galaxy.

\subsection{Large rotating \HI\ structures }

About two-thirds of the galaxies from our sample have a more or less
regular rotating \HI\ structure.

The observed \HI\ disks are sometimes very large. Their size is indicated in
Tables 2 and 4 and ranges from a few tens of kpc up to more than 200 kpc in
the case of NGC~6799. The kinematics of the \HI\ structures suggests that the
gas in these large structures has fairly regular rotation -- at least at the
low spatial resolution of our observations. However, it is clear that the low
resolution and limited depth of the data are not sufficient to investigate the
structures in all detail.  Some of the cases here classified as disks may in
fact be strongly warped or even polar rings. For example, in ESO~140-31 the
\HI\ structure appears to be perpendicular to the optical major axis. At the
same time, the kinematics of NGC~6799 and IC~4889 suggests that a  warp of
$\sim$90$^\circ$ is present, similar to the already known cases such as 
NGC 1947, NGC 3108 (Oosterloo et al.\ 2002) and NGC~5266
(Morganti et al.\ 1997). Recent higher resolution and deeper observations of IC 4200
(Serra et al.\ 2006) show that also in this galaxy the disk is strongly
warped.  More elaborate observations of several galaxies discussed in this
paper are already in progress to investigate the \HI\ structures of some of
these disks in more detail.

\subsection{Tidal tails and offset clouds}

There are 10 galaxies in our sample  for which the \HI\ is distributed in tail-like
structures or in clouds offset from the target galaxy.  The presence  of
these tails and clouds suggest that interactions occurred where some gas is
captured from a companion.

An intriguing example is the field elliptical galaxy NGC\,1490. A large
\HI\ cloud, together with 3 smaller \HI\ clouds, forms an arc of $\sim 500$
kpc long that half encircles NGC\,1490 (see Fig.\ 2 and Oosterloo
et al.\ 2004 for more details). The total \HI\ mass of the \HI\ clouds is $8
\times 10^{9}$ $M_\odot$. None of the clouds has an optical counterpart on DSS
images, but deeper images show that the largest cloud has a faint, low surface
brightness optical counterpart. This is the \HI-richest object known with an
optical counterpart: $M_{\rm \HI}/L_{\rm V}$ $\sim 165\
M_\odot/L_{\odot,V}$. Most interestingly, two small \HII\ regions are present
at the edge of the largest \HI\ cloud, both about 80 kpc removed from any
object detected in deep broadband optical images, and they appear to be
inter-galactic \HII\ regions.  The H$\alpha$ luminosities of the \HII\
complexes are $1-3 \times 10^{38}$ erg s$^{-1}$ (i.e.\ 25-75 times Orion) and
the ionised gas has metalicities around 0.25-0.5 solar (Oosterloo et al.\
2004). NGC 1490 does not show any signs of accretion or interaction (A.\
Ferguson priv.\ comm.). The size of the system of \HI\ clouds and the
relative velocities imply a (kinematical) timescale of about 1 Gyr. This is
too short for the signatures of a significant merger to disappear. The \HI\
clouds found in the field of NGC~1490 are therefore most likely remnants of a
galaxy that has been torn apart in a tidal interaction with NGC 1490 - tidal
shredding - while passing a larger galaxy, without merging or accretion taking
place (e.g.\ Bekki et al.\ 2004) . This kind of interaction can lead to
structures of very low surface brightness well outside a large, apparently
undisturbed galaxy.

\begin{figure*}
\centerline{\psfig{figure=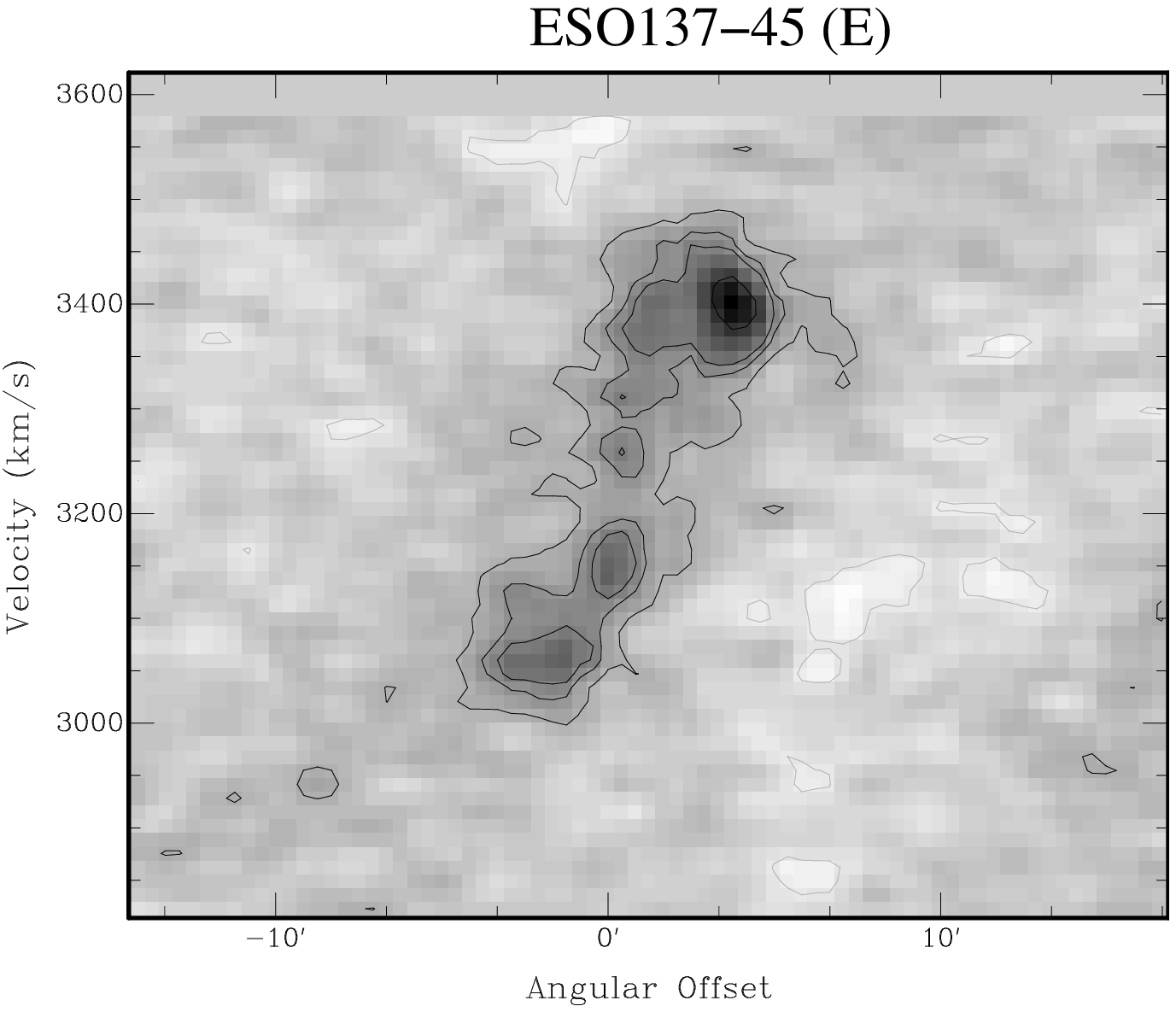,angle=0,width=7cm}
\hskip1cm        \psfig{figure=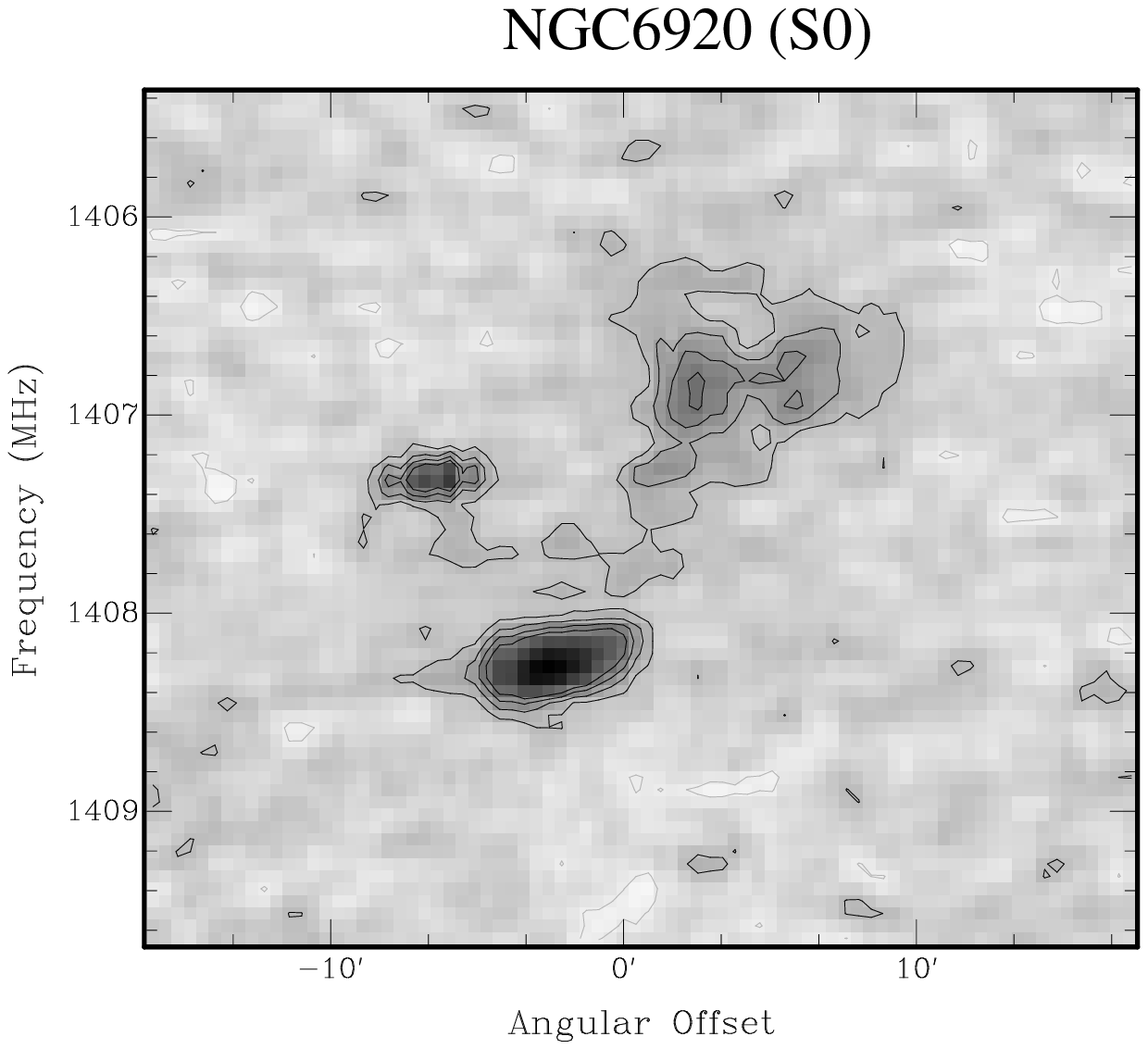,angle=0,width=7cm}}

\centerline{      \psfig{figure=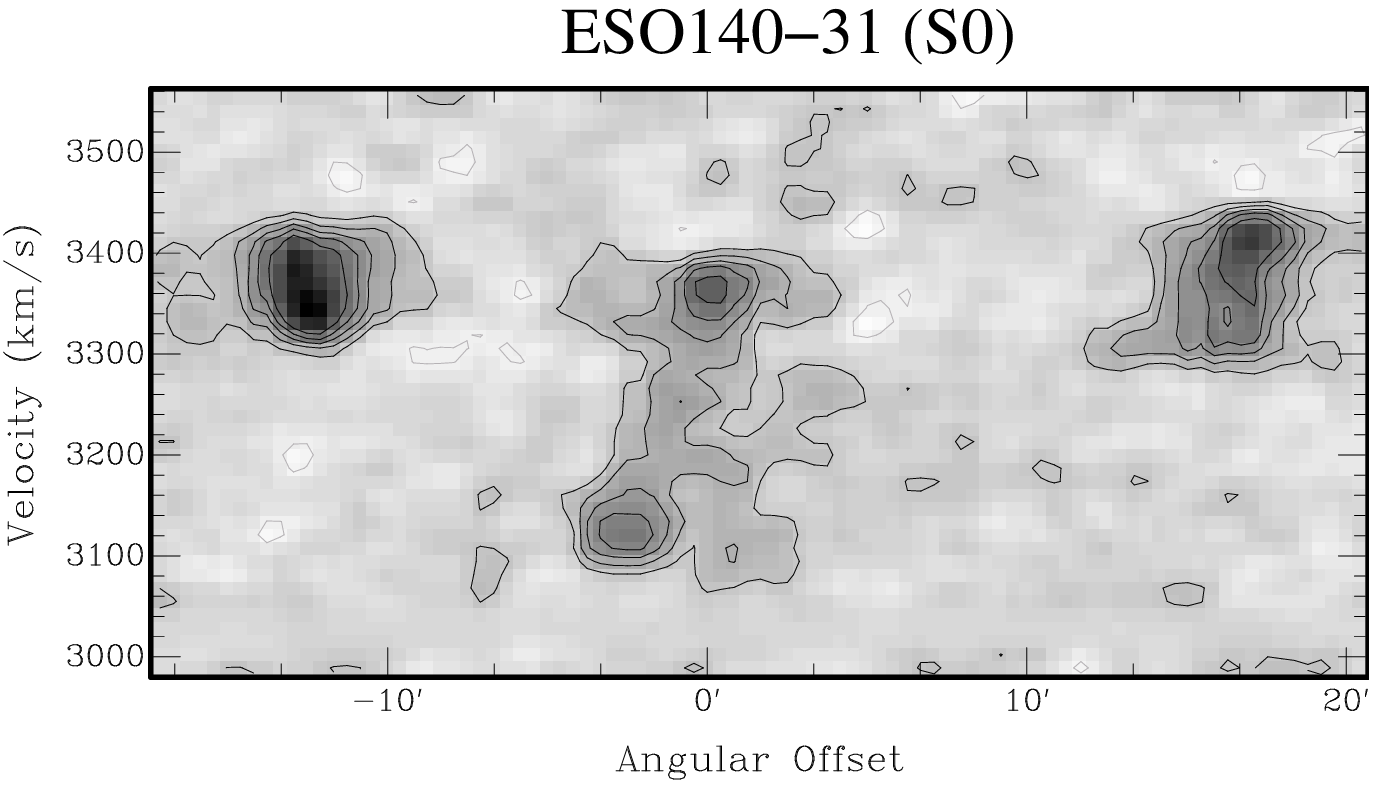,angle=0,width=7cm}}

            \caption{Position-velocity plots along the major axis for three of
            the largest \HI\ disks observed in the sample. ESO~137--45,
            ESO~140--31 (together with two companions) and NGC~6920 (together
            with one companion).  Contour
            levels are: --3, 3, 6, 8, 16 mJy beam$^{-1}$ in the case of
            ESO~137--45, --3, 3, 6, 9, 12 mJy beam$^{-1}$ for ESO~140--31 and --3,
            3, 6, 8 mJy beam$^{-1}$ for NGC~6920.} \end{figure*}

\subsection{Confusion \& Environment}

One of the motivations for the follow up ATCA observations was to quantify the
effect of confusion on the detection statistics due to the large beam of the
original HIPASS survey.  The effects of confusion will be discussed in more
detail in Paper II.  However, from our observations it
is clear that about half of the early-type galaxies whose HIPASS spectra show
\HI\ emission, have a nearby companion (i.e.\ inside the $\sim 15$ arcmin
Parkes beam) with similar velocity, underlining that confusion is a major
factor in studying the statistical properties of \HI\ based on single-dish
data such as HIPASS. The fact that confusion is an important factor also for
early-type galaxies does show that many  early-type galaxies do reside in a
\HI-rich environment.

\section{Discussion}
\label{discussion}

\subsection{\HI-rich early-type galaxies}

The most important result from the observations presented here is the finding
that in many \HI-rich early-type galaxies the neutral hydrogen is distributed
in very large, low surface brightness disk-like structures with large \HI\
masses.

The existence of  such structures was already known for a few  single
cases (Morganti et al.\ 1997, 1998, Oosterloo et al.\ 2002, van Gorkom \&
Schiminovich 1997), while also the much deeper observations of the small sample
of \sauron\ early-type galaxies indicated that such disk-like structures exist.  The
present study is important because it indicates that these massive disk-like
gaseous systems are relatively common in gas-rich early-type galaxies. This makes them
relevant for the theories that describe the formation and evolution of E/S0's.

Interestingly, as can be seen from Fig.\ 5, there is some difference
between the \HI\ morphologies properties of elliptical and of S0
galaxies. Tails and offset clouds of \HI\ appear to be somewhat more common in
elliptical galaxies, although the difference is statistically speaking not
very significant ($50\pm22\%$ vs $25\pm11\%$).  We note that the large,
regular structures occur in both types of galaxies, indicating that such
structures are not strictly connected with the presence of a stellar
disk. This confirms what was found for the
\sauron\ sample (Morganti et al.\ 2006).  These authors do not find evidence
for a clear trend of the presence of \HI\ with either the global age of the
stellar population or with the global dynamical characteristics of the
galaxies.  More specifically, for the
\sauron\ galaxies, the \HI\ detections are uniformly spread through the
$(V/\sigma,\epsilon)$ diagram. If fast and slow rotators - galaxies with
high and low specific angular momentum - represent the relics of
different formation paths, this is not obvious from  the presence and
characteristics of the \HI.
 
Fig.\ 3 shows the distribution of $M_{\rm HI}/L_B$. Unlike what observed in
spirals, but similar to what earlier observed for early-type galaxies by Knapp
et al.\ (1985), the distribution appears very broad.  This is especially clear
if we take into account also the distribution found in deep \HI\ observations,
i.e.\ observations more sensitive than HIPASS.  This distribution has led
previous authors to the conclusion that in early-type galaxies the gas and the
stellar content are decoupled and that in most early-type galaxies the gas has
an external origin. However, recent studies of nearby spiral galaxies,
including our own, show that also these galaxies are continuing to accrete
material (van der Hulst \& Sancisi 2004; Westmeier, Braun \& Thilker
2005). The lack of correlation between the optical luminosity and the gas
content observed for early-type galaxies must therefore have a more complex
explanation. The very low column density of the gas disks observed in
early-type galaxies could imply that what distinguishes a gas-rich early-type
galaxy from a spiral is not so much whether accretion occurs or not, but
instead is what happens with the material after it has been accreted.  The
observed column density of the
\HI\ typically peaks at values around 
$\sim$10$^{20}$ cm$^{-2}$ (the maximum column density is $3.2 \times 10^{20}$
cm$^{-2}$, observed in ESO~92--21). This is expected to be below the critical
density for star formation. Given the low resolution of our data, it is likely
that locally the density will be above the threshold for star formation and
that stars are forming in a few small regions. However, star formation will not occur
on any large scale.  This implies that the disks will only slowly be consumed
by star formation and that despite the large \HI\ reservoir, the gas disks can
stay intact for very long periods of time and that large stellar disks will
develop only slowly over time. Perhaps galaxies with such a large \HI\
structure are related to giant LSB galaxies like Malin 1 (Impey \& Bothun,
1989). Such galaxies are characterised by a bright bulge surrounded by a large, very
low surface brightness optical disk. If such a relation exist, deep optical
imaging of the galaxies of our sample could reveal such LSB optical disks.

The large \HI\ disks appear to have relatively regular kinematics. Although
this will need to be further investigated with much deeper observations of
higher resolution, the overall regularity of the kinematics is indicating that
the gas is settled and therefore these disks must be relatively old and
long-lived. The gas must have been around long enough in order to settle into
a regular configuration. Thus, whatever is the process that created these
structures, it must have happened at least a few times $10^9$ years ago
(corresponding to a few orbital periods in the outer regions of the disks),
with the oldest disks having formed before $z=0.5$. These large gas disk must
be relics of an important aspect of the evolution of these early-type
galaxies.

\begin{figure}
\centerline{\psfig{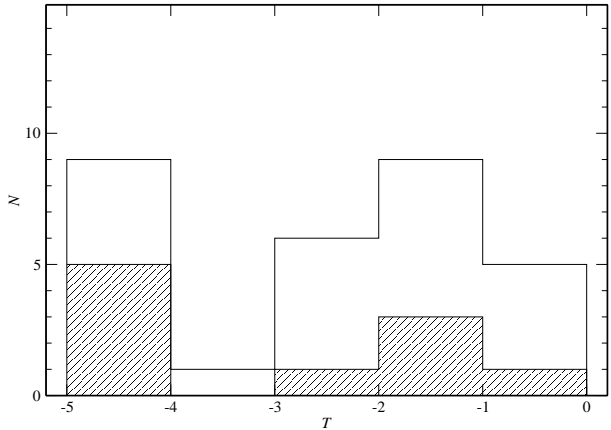}}
\caption{Distribution of optical morphological type of the \HI\
detections. The shaded area indicates the fraction of irregular \HI\ morphologies}
\end{figure}

\subsection{Origin of the \HI\ disks}
\label{discorigin}

As mentioned in the Introduction, the observed gas-rich galaxies can form via
major merger, with the \HI\ disks originating from the late-infall of
high-angular-momentum gas. Simulations (Hibbard \& van Gorkom 1996; Hibbard \&
Mihos 1995, Barnes 2002, Naab et al.\ 2006) show indeed that at first this gas
forms the extended tidal tails and bridges observed in many merging
systems. Later, it can be re-accreted by the newly formed galaxy and settle
into an extended but dilute rotationally-supported disk. Indeed, some of the
\HI-rich early-type galaxies already known to be gas rich have 
been interpreted as the result of a major
merger of gas-rich galaxies (e.g., Centaurus A, NGC~5266 and
NGC~3108). Scaling the results of the simulations to the stellar mass of the
sample galaxies, one can reproduce gas-disks of size and mass comparable with
those observed.  The presence of large amount of \HI\ in regular structures
could be connected to a specific type of merger. For example, Burkert \& Naab
(2004) suggest that the extended disk-like component naturally form in
gas-rich fast-rotating, 3:1 merger remnants (Naab \& Burkert 2001) although
even in 1:1 mergers the remaining gas in the outer parts of the remnant has
high enough angular momentum to form extended gas disks as it falls back
(Barnes 2002).

If the neutral hydrogen in early-type galaxies is acquired via merger and/or
accretion, one would perhaps have expected a larger number of tail-like
structures or offset \HI\ structures in our sample.  The relatively small
number of such structures is possibly due to the fact that our survey is
shallow and that these tails may tend to have smaller masses. An indication
for this comes from more sensitive observations of the \sauron\ sample
(Morganti et al.\ 2006) where the fraction of galaxies with tails and offset
clouds is somewhat higher. It is possible that a very extended and regular gas
configuration is typical only of systems with a large \HI\ mass, while
irregular morphologies are easier to find among the less massive ones.  An
other possibility is that the rate of major mergers have been higher in the
past. Indeed, given the size and regular appearance of the observed \HI\
disks, they must be quite old, in many cases well over $5
\times 10^9$ yr. 
%
%Certainly, as the results of Morganti et al.\ (2006) confirm, the
%presence of \HI-gas in field early-type galaxies is starting to appear as a
%common characteristic of these objects and the presence of regular disc-like
%structures appears to be very common.

An alternative to the merger hypothesis is that accretion of gas from the IGM
would also result in the formation of extended and regular cold-gas structures
with masses up to 10$^{10}$\msun.  According to simulations, a fraction of the
accreted gas is not shock-heated to the virial temperature of the accreting
halo, but is kept below $\sim$10$^5$ K (Binney 2004, Keres et al.\ 2005 and
references therein). This colder gas can cool over a reasonable time-scale
(provided the environment is not too hostile so it is not destroyed by
neighbouring galaxies, and a few-percent enrichment from the stellar processes
in the galaxy is allowed) providing the necessary supply of atomic gas. In
some simulations, indeed large gaseous disks are found around galaxies that
also show kinematics similar to that observed here, e.g., in the form of polar
disks (Macci\'o et al. 2006).  So it appears that the formation of
\HI\ disks via cold accretion is viable.

A possible way to distinguish between the two scenarios is by studying
different properties of the \HI-rich galaxies in order to find other
signatures of their evolution. If the formation of the gas disk can be traced
back to a discrete event, than a merger seems a more likely explanation for
its origin, while if it is the result of smooth accretion from the IGM, no
such signature would be expected in, for example, the stellar
population. Serra et al.\ (2006) showed this to be a good approach by relating
deep \HI\ observations, stellar populations analysis and the study of optical
morphology for one of the galaxies of our sample (IC~4200). They found a
strongly warped \HI\ disk (containing almost $10^{10}\ M_\odot$ of \HI), a
young stellar population in the centre of the galaxy and stellar
shells. Furthermore, they could match the timescale for the formation of the
gas disk, as estimated from the \HI\ morphology/kinematics, with the presence
of a distinct young stellar population. They conclude that IC~4200 formed via
a major merger of two Milky-Way-like galaxies between 1 and 3 Gyr ago. These
authors are carrying out a similar analysis on a sample of galaxies that
contains several objects presented in the present paper and this will allow to
make a more statistical assessment. It is clear, however, that the gas
accretion cannot always be directly connected to the stellar population.
Morganti et al.\ (2006) find that the properties of the stellar population of
 early-type galaxies do
not correlate with the gas content, i.e., galaxies with a purely old stellar
population as well as those where a young sub-population is present, can be
either gas-poor or gas-rich. This suggests that at least in some cases
accretion from the IGM may have occurred. Their sample mostly contains
galaxies that have (much) less \HI\ than those of the sample presented here
and it remains to be seen whether the correlation is also absent for
early-type galaxies containing more \HI.

\begin{figure}
\centerline{\psfig{figure=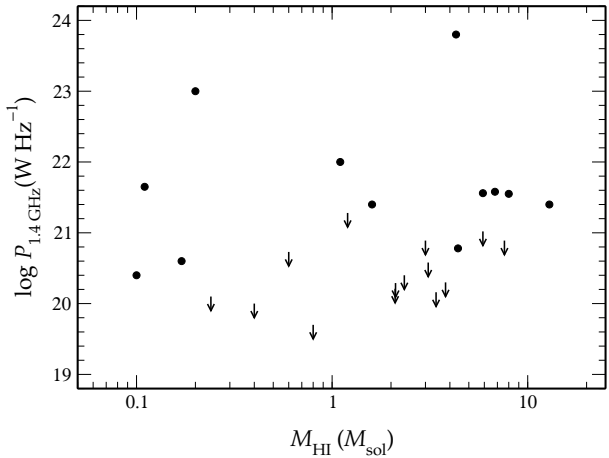,width=8cm}}
\caption{Radio power plotted vs \HI\ mass for the galaxies detected in HI}
\end{figure}

\subsection {(Radio-loud) AGN and neutral hydrogen}

Nuclear activity is often claimed to be related to the presence of interaction
or merger in the life of a galaxy.  Torques and shocks during the merger can
remove angular momentum from the gas in the merging galaxies and this provides
injection of substantial amounts of gas/dust into the central nuclear regions
(see e.g.\ Mihos \& Hernquist 1996).  Because of this, it is of interest to
explore whether any relation exists between the presence of continuum emission
and the presence/structure of the neutral hydrogen.

The radio continuum for each target (or an upper limit to it) has been derived
using the line-free channels of our ATCA observations.  The fluxes and
corresponding radio powers are listed in Table 2 and 4, and are plotted vs the
\HI\ mass in Fig.\ 6.  It is clear that there is no correlation between the
radio continuum and the \HI\ mass.

Although our data cannot tell us whether the continuum emission is originating
from star formation or from the active nucleus, emission from the latter is
clearly present at least in some cases (e.g.\ IC 4200 and IC 5063).  A
detailed study of the origin of the continuum emission will need supplementary
data. However, it is interesting at this stage to point out that the
large-scale \HI\ structures, that are possibly originating from major mergers,
do not necessarily host a radio-loud source and none of them host a {\sl
large-scale} radio galaxy. This is interesting as gas-rich mergers between
disk galaxies have been claimed to be at the origin of powerful radio
galaxies. As describe above, these type of merger can form the large, \HI-rich
disks detected in this survey.  The lack of detection of large radio galaxies
could be related to some bias in the selection inside the HIPASS survey:
sources with too strong a radio continuum would not be easily detected in \HI\
using single dish observations. However, it would be consistent with what
obtained from a \HI\ survey in nearby radio galaxies (Emonts et al.
2006). From this study, the largest ($\sim 100$ kpc scale) and richest in
\HI\ ($>10^9 M_\odot$) structures are only associated to {\sl compact} radio
sources. These are sources that unlikely will grow to become large radio
galaxies but may become comparable (once that the main phase of radio activity
is past) to the galaxies with radio continuum and large \HI\ structures
observed in our HIPASS survey.  Thus, all this indicates that either the
trigger and continuous fuel needed to produce a large-scale radio galaxy
happens through a different mechanism than the one producing the large
\HI\ disks or, alternatively,  the presence of an extended radio galaxy
prevents the formation of large \HI\ structures (see also Emonts et al.\ 2006
for further discussion).

\section{Conclusions}
\label{conclusions}

In this paper we have discussed the morphology and kinematics of the \HI\ of 30 gas-rich early-type galaxies homogeneously selected on the basis
of their \HI\ properties. Given that these galaxies were selected from the
HIPASS, they correspond to the most gas-rich early-type galaxies. The main
result is that in about 2/3 of the galaxies the \HI\ is in a regularly
rotating disk- or ring-like structure. Only in about 1/3 of the galaxies the
\HI\ shows irregular morphologies. The high incidence of regular \HI\ structures
found in this sample confirms what was found for the \sauron\ sample of
early-type galaxies (Morganti et al.\ 2006) for which a large fraction of
regular \HI\ structures is also detected. The regular disk-like structures
shown here can be up to 200 kpc in size and contain up to $10^{10}$ $M_\odot$
of \HI.  Several of these disks are strongly warped. Given their size,
some of these disks must be old and must have formed several Gyr ago. The
column density of the \HI\ is low and  only very little star formation is
expected to occur. Perhaps these systems are related to giant LSB galaxies
like Malin 1. Given the low star formations rate, these disks will remain
intact for very long periods of time.  The
\HI\ properties  correlate only weakly with the optical classification and
\HI\ disks are detected around both E and S0 galaxies, as is also the case for
the galaxies from the \sauron\ sample.  Therefore,  if these two
types of galaxies are the result of different formation paths, this is not strongly
reflected in the presence and characteristics of the \HI. Given the current
data, both merging and gas accretion from the IGM are viable explanations for
the origin of the gas in these galaxies. A more detailed investigation of the
correlation with, e.g., the properties of the stellar population, is required
in order to be able to distinguish between these two mechanisms.

\begin{acknowledgements}
This work is based on observations with the Australia Telescope Compact Array
(ATCA), which is operated by the CSIRO Australia Telescope National
Facility. This research has made use of the NASA Extragalactic Database (NED),
whose contributions to this paper are gratefully acknowledged.  The Digitized
Sky Survey was produced at the Space Telescope Science Institute under US
Government grant NAG W-2166
RM and TO wish to thank the University of Sydney for the support during their 
visits.

\end{acknowledgements}

\newpage

\begin{table*} 
\centering 
\caption{Parameters of the ATCA observations for the E and S0 galaxies
selected from the HIPASS south of dec -30 candidates \HI\ detections.
Noise for natural weighting images.}
\begin{tabular}{lcclrrccccc} 
\hline\hline\\ 
Name         &  RA        &  Dec       & Type & $T$  & K     & IRAS $S_{60}$ & 
$V_{\rm opt}$& Obs.   & Exp.  & noise \\ 
             & J2000.0    & J2000.0    &      &    & mag.  &  Jy         & 
\kms\    &        & h     &    \mJybeam\    \\
\hline \\ 
NGC\,1490    & 03:53:34.5 &$-$66:01:02 & E1  & --5.0&  9.05 & $<$0.033 & 5397    
& 20Oct01& 12   & 1.0    \\
ESO\,88-18   & 07:29:06.1 &$-$66:54:44 & E5P & --5.0& 11.72 & ...      & 5055    
& 22Oct01&  2.1 & 3.6  \\
NGC\,2434    & 07:34:51.5 &$-$69:17:01 & E0  & --5.0&  7.85 & $<$0.023 & 1390    
& 22Oct01&  2.1 & 3.5   \\
ESO\,318-21  & 10:53:05.3 &$-$40:19:46 & E?  & ... &  9.52 & $<$0.044 & 4831    
& 15Mar01&  8.1 & 1.4   \\ % 6dFGS spectrum 
ESO\,137-45  & 16:51:03.5 &$-$60:48:30 & E0  & --5.0&  8.54 & 0.180    & 3351
& 07Mar02& 13   & 1.3   \\ %6dFGS spectrum 
ESO\,138-1   & 16:51:20.6 &$-$59:14:11 & E?  & ... & (9.75)& 2.648    & 2740    
& 09Mar02& 10.6 & 1.2    \\ % Gal. lat. -9.4 deg 
ESO\,140-31  & 18:37:54.0 &$-$57:36:42 & E1  & --5.0&  9.50 & ...      & 3150    
& 03Jul01&  5.7 & 1.5  \\ 
IC\,4765     & 18:47:18.5 &$-$63:19:57 & E4  & --4.0&  8.11 & 0.130    & 4507    
& 02Jul01&  1.5 & 3.2  \\ 
IC\,4889     & 19:45:15.9 &$-$54:20:37 & E5  & --5.0&  8.00 & 0.160    & 2574    
& 03Jul01&  5.7 & 1.5   \\ % 6dFGS spectrum 
NGC\,6877    & 20:18:37.1 &$-$70:51:16 & E6  & --5.0& (...) & ...      & 4440    
& 12Jun01&  2.9 & 2.3   \\ %in IC\,4970 field??  
 & & & & & & & & & &  \\
IC\,1724     & 01:43:09.1 &$-$34:14:31 & S0  & --1.0& 10.02 & ...      & 3816    
& 21Oct01&  1.3 & 3.3    \\
ESO\,153-4   & 01:58:30.5 &$-$56:14:57 & S0  & --3.0& (...) & 0.390    & 5936    
& 22Oct01&  1.5 & 3.3    \\
IC\,1813     & 02:30:49.1 &$-$34:13:14 & S0  & --0.5& 10.04 & 0.220    & 4453    
& 21Oct01&  1.2 & 3.8   \\ % NVSS source, 3.6 mJy, 6dF spectrum 
IC\,2041     & 04:12:34.3 &$-$32:48:59 & S0  & --2.0& 11.65 & ...      & 1260    
& 21Oct01&  1.5 & 3.3   \\ % 2dFGRS S922Z098, z=0.003680 
IC\,2122     & 05:19:01.1 &$-$37:05:16 & S0  & --2.7&  9.40 & ...      & 4663    
& 21Oct01&  2.0 & 3.1   \\
IC\,2200A    & 07:28:06.7 &$-$62:21:45 & S0  & --2.7&  9.47 & ...      & 3248    
& 22Oct01&  2.3 & 3.3   \\
NGC\,2904    & 09:30:16.9 &$-$30:23:01 & S0? & --3.0&  9.19 & $<$0.075 & 2395    
& 21Oct01&  2.7 & 4.0   \\  % 6dFGS spectrum 
ESO\, 92-21  & 10:21:05.5 &$-$66:29:31 & S0  & --3.0& (9.76)& ...      & [2023]  
& 14Mar02&  2.5 & 3.1  \\ % in HIPASS brightest 1000, gal. lat. -7.9 deg 
ESO\,215-32  & 11:11:22.9 &$-$48:01:07 & S0? & --2.0&  9.88 & ...      & 4311    
& 08Mar02&  2.1 & 2.8   \\ 
MCG-7-26-21  & 12:39:50.4 &$-$41:06:20 & S0  & --2.0&  9.46 & ...      & 4391    
& 08Mar02&  2.1 & 2.9   \\ % 6dFGS spectrum  
ESO\,322-84  & 12:48:00.0 &$-$39:37:51 & S0  & --2.0& 10.26 & ...      & 4469    
& 08Mar02&  2.4 & 2.8   \\ % 6dFGS spectrum  
ESO\,323-13  & 12:51:32.4 &$-$41:13:42 & S0? & ... & 10.44 & 0.374    & 4802    
& 08Mar02&  2.1 & 3.0   \\ % 6dFGS spectrum, NED classification is Sb!! 
ESO\,269-12  & 12:56:40.8 &$-$46:55:31 & S0? & --2.0\rlap{$^a$} & 10.13 & ...      & 
4950    & 10Mar02&  2.1 & 3.6   \\ % Listed as Sy2 in NED
ESO\,269-14  & 12:57:06.9 &$-$46:52:19 & S0? & ... & 10.82 & 1.556    & 4850    
& 10Mar02&  2.1 & 3.6   \\ % field ESO\,269-12, 16.7 mJy in SUMSS  
ESO\,381-47  & 13:01:05.4 &$-$35:37:00 & S0? & --2.0\rlap{$^a$} & 10.50 & ...      & 
4771    & 08Mar02&  2.0 & 3.2   \\ % 6dFGS spectrum  
IC\,4200     & 13:09:34.7 &$-$51:58:07 & S0  & --2.0&  8.72 & ...      & 3938    
& 10Mar02&  1.8 & 3.1   \\
NGC\,4988    & 13:09:54.1 &$-$43:06:22 & S0  & --1.4& 10.23 & 1.529    & 2097    
& 14Mar02&  2.0 & 2.7   \\ 
ESO\,269-80  & 13:19:01.7 &$-$47:15:27 & S0  & --3.0&  9.01 & 0.550    & 3202    
& 15Mar02&  1.3 & 3.3    \\
NGC\,5234    & 13:37:29.9 &$-$49:50:16 & S0  & --0.6&  9.51 & ...      & 3650    
& 14Mar02&  2.2 & 2.8   \\ 
IC\,4312     & 13:40:31.3 &$-$51:04:16 & S0  & --2.0&  8.81 & 2.106    & 4001    
& 10Mar02&  1.7 & 3.2   \\ 
ESO\,221-20  & 13:58:23.4 &$-$48:28:29 & S0p & --3.0&  9.24 & ...      & 2780    
& 14Mar02&  2.4 & 2.7   \\ 
ESO\,137-44A & 16:50:56.0 &$-$61:48:54 & S0? & --2.0&  8.60 & ...      & 4493    
& 10Mar02&  1.9 & 3.8   \\ % 6dFGS spectrum 
IC\,4647     & 17:26:02.2 &$-$80:11:41 & S0  & --2.0& 10.24 & ...      & 4872    
& 10Mar02&  1.9 & 4.0   \\ 
ESO\,103-49  & 18:42:02.8 &$-$65:05:50 & S0? & ... & 10.37 & ...      & 4610
& 12Jun01&  2.9 & 2.3  \\ % 6dFGS spectrum 
IC\,4847     & 19:23:32.7 &$-$65:30:25 & S0? & --2.0\rlap{$^a$} & 11.35 & 0.378    & 
4300    & 11Jun01&  2.5 & 1.5\rlap{$^{*}$}  \\
             &            &            &     &     &       &          &         
& 02Jul01&  4.3 &       \\ 
NGC\,6799    & 19:32:16.2 &$-$55:54:30 & S0  & --3.7&  9.46 & $<$0.050 & 5990?   
& 11Jun01&  2.5 & 1.0   \\ 
             &            &            &     &     &       &          &         
& 01Jul01& 12   &       \\          
             &            &            &     &     &       &          &         
& 04Jul01&  5.4 &       \\
NGC\,6850    & 20:03:29.9 &$-$54:50:45 & S0p & --1.0&  9.19 & 0.426    & 4950    
& 12Jun01&  3.0 & 1.3   \\ % 6dFGS spectrum 
             &            &            &     &     &       &          &         
& 02Jul01&  3.9 &       \\ 
IC\,4970     & 20:16:57.9 &$-$70:44:57 & S0p & --3.2& 10.89 & ...      & 4727    
& 12Jun01&  2.9 & 2.8   \\ % small companion to spiral NGC 6872
ESO\,234-11  & 20:22:02.7 &$-$47:59:12 & S0  & --1.6& 10.38 & ...      & 5702    
& 12Jun01&  2.7 & 2.4   \\ % 6dFGS spectrum 
NGC\,6920    & 20:43:56.6 &$-$80:00:03 & S0  & --2.0&  8.34 & 0.310    & 2774    
& 11Jun01&  2.2 & 2.2   \\ % in HIPASS cat. 
             &            &            &     &     &       &          &         
& 04Jul01&  4.9 &       \\ 
NGC\,7166    & 22:00:32.8 &$-$43:23:22 & S0  & --3.0&  8.48 & 0.210    & 2426    
& 11Jun01&  2.0 &       \\ % 6dFGS spectrum 
ESO\,240-10  & 23:37:44.3 &$-$47:30:17 & S0p & --2.0&  8.69 & ...      & 3221    
& 11Jun01&  1.7 & 2.8   \\ % 6dFGS spectrum 
 & & & & & & & & & &  \\
\hline\hline
\end{tabular}
\\
Notes: 
$^{*}$ when more than one observations is
available, the noise is from the combined data;
$^a$ From ESO Cat on NED.

\end{table*}
\newpage

%\begin{landscape} 
\begin{table*} 
%\begin{center}
%\centering
\caption{Observed  parameters for the  sources. Upper limits to the continuum
flux are $3\sigma$.}
\begin{tabular}{lrcrccrrl} 
\hline\hline\\ 
Name &  $L_B$ & $S_{\HI}$ & $M_{\HI}$  & $\Delta V(20\%)$ &  $M_{\HI}$/$L_B$
&  $S_{\rm 1.4\ GHz}$ &
log $P_{\rm 1.4\ GHz}$ & Morphology   \\ 
     & $10^9$$L_\odot$ & Jy km s$^{-1}$ & $10^9$ $M_\odot$ & \kms & &  ( mJy &
W Hz$^{-1}$ &Contour levels (CL) in   cm$^{-2}$ \\
\hline
Ellipticals\\
\\
NGC\,1490    & 34.9  &  5.7 & 8.0    & ... &  0.23 & 5.7    & 21.55     & \HI\ 
tail/ring $\sim$ 100 kpc  \\
             &       &     &    &     &      &        &           &  from the 
galaxy$^1$  \\
%\hline
ESO\,88-18   & 9.0   &  ... & $<1.4$  & ... & $<0.15$ & 20.0   & 22.09     & \\
%\hline
NGC\,2434    & 4.3   &  ... & $<0.1$  & ... & $<0.02$ & $<7.5$ & $<20.36$  & \\
%\hline
ESO\,318-21  & 17.9  & ...  & $<0.6$  & ... & $<0.03$ & $<1.3$ & $<20.81$  & \\
%\hline
ESO\,137-45  & 14.5  & 13.34  & 6.8    & 480 & 0.47 & 16.8   & 21.58     & disk of
$\sim 110$ kpc  \\
             &       &      &        &     &      &        &           & 
CL: 4, 8, ... $\times 10^{19}$ \\
%\hline
ESO\,138-1   & 3.5   & 0.31  & 0.11    & ...    &  0.03  & 30.6   & 21.65     &
possible \HI\ off-galaxy cloud or \\
             &       &      &       &     &      &        &           & companion 
$\sim 70$ kpc from target\\
             &       &      &      &     &      &        &           &  
2, 4, 8, 16, ... $\times 10^{19}$  \\
%\hline
ESO\,140-31  & 9.1   &  7.7 &  3.8    & 310 & 0.42 & $<1.0$ & $<20.30$  & disk or
polar ring($\sim 76$ kpc) \\
             &       &      &   &     &      &        &           & 
CL: 2, 4, 8, 16, ... $\times 10^{19}$ \\ 
%\hline
IC\,4765     & 68.6  & ...  & $<1.0$  & ... & $<0.01$ &  5.6   & 21.41     &    \\ 
%\hline
IC\,4889     & 20.6  & 7.1  &  2.1    & 160 & 0.10 & $<1.3$ & $<20.21$  & disk of
$\sim 80$ kpc in size \\
             &       &      &   &     &      &        &           & 
CL: 2, 4, 8, 16, ... $\times 10^{19}$  \\ 
%\hline
NGC\,6877    & 28.7  & ...  & $<0.9$  & ... & $<0.03$ & $<1.2$ & $<20.69$  & IC~4070
field \\ 
             &       &      &         &     &      &        &           & \\
\hline
%\hline
S0s\\
             &       &      &         &     &      &        &           & \\
IC\,1724     & 9.2   & ...  & $<0.73$  & ...   & $<0.08$ & $<1.2$ &  $<20.56$         &   
\\
%\hline
ESO\,153-4   & 14.1  & ...  & $<1.7$  & ... & $<0.12$ & $<1.7$ &  $<21.11$         &  \\
%\hline
IC\,1813     & 12.1  & ...  &  $<0.76$    & 125 & $<0.06$ & $3.6^*$ &    21.20      &
interacting? \HI\ tail\\
             &       &      &    &     &      &        &           &  ($\sim 110$ 
kpc) from a companion \\
             &       &      &    &     &      &        &           & 
CL: 4, 8, 16, ... $\times 10^{19}$ \\
%\hline
IC\,2041     & 0.6   &  ... & $<0.1$  & ... & $<0.18$ & $0.9$  &   19.44        & \\
%\hline
IC\,2122     & 18.4  &  5.7 & 5.9    & 200? &0.32 & $<2.2$ &    $<21.02$       & disk 
(of $\sim 74$ kpc), interacting? \\
             &       &      &   &     &      &        &           & 
CL: 4, 8, 16, ... $\times 10^{19}$ \\
%\hline
IC\,2200A    & 8.8   & ...  & $<0.5$  & ... & $<0.06$ & 15.4&   21.51    & 
dumbell system \\
 &    &   &   &  &  & $<0.9$&   $<20.28$   \\
%\hline
NGC\,2904    & 5.1   & ...  & $<0.4$  & ... & $<0.08$ & $<1.0$ &   $<20.06$  &  \\
%\hline
ESO\, 92-21  & 2.7   & 19.2 & 3.4     & 250 & 1.26 & $<2.0$ &   $<20.16$      & disk 
($\sim 57$ kpc in size)\\
             &       &      &    &     &      &        &           & 
CL: 2, 4, 8, 16, ... $\times 10^{19}$\\
%\hline
ESO\,215-32  & 10.7  & 0.68  & 0.6  & ... & 0.06 & $<1.4$ &   $<20.73$        & off-
galaxy cloud, \\
             &       &      &    &     &      &        &           & about $\sim 
150$ kpc from target \\ 
             &       &      &   &     &      &        &           & 
CL:  2, 4, 8, 16, ... 10$^{19}$ \\
%\hline
MCG-7-26-21  &       & ...  & $<0.9$  &  ...   & ...   & $<2.0$     & $<20.60$          & \\ 
%\hline
ESO\,322-84   & 9.4  & ...  & $<0.9$  & ... & $<0.10$ & 15.26  &      21.81 &      \\ 
%\hline
ESO\,323-13   & 10.3 & ...  & $<1.0$  & ... & $<0.10$ &   $<1.1$     &    $<20.70$       & \\
%\hline
ESO\,269-12   & 8.9  & 1.0  & 1.2     &     & 0.13 & $<3.7$ &       $<21.28$    & faint 
disk ($\sim 60$ kpc) or tail \\ 
             &       &      &      &     &      &        &           & toward 
ESO\,269-14\\ 
             &       &      &    &     &      &        &           &
CL: 2, 4, 8, 16, ... $\times 10^{19}$ \\
%\hline
ESO\,269-14   & 9.7  & ...  & $<1.3$  & ... & $<0.13$ &  9.0   &  21.65         & see 
ESO\,269-12\\
%\hline
ESO\, 381-47  & 15.9 & 7.1 & 7.6     & 130 & 0.48 & $<1.2$ &   $<20.76$    & disk 
($\sim
140$ kpc) \\
             &       &      &   &     &      &        &           &+ cloud ($\sim 
250$ kpc from target) \\
             &       &      &   &     &      &        &           & 
CL:  2, 4, 8, 16, ... $\times 10^{19}$ \\
%\hline
IC\, 4200     & 11.7 & 8.7 & 5.9    & 210 & 0.51 & 11.5   &    21.56       & disk$^2$
(about 160 kpc in size) \\
             &       &     &    &     &      &        &           & 
CL: 2, 4, 8, 16, ... 10$^{19}$ \\
%\hline
NGC\,4988     & 3.1  & 10.0 & 2.11    & 170 & 0.68 & $<2.4$ &   $<20.29$     & disk
of $\sim 68$ kpc  \\ 
             &       &     &    &     &      &        &           &
CL: 4, 8, 16, ... $\times 10^{19} $ \\
%\hline
ESO\,269-80   & 7.9  & ... & $<0.51$      &  ...   & $<0.06$   & $<2.0$    & $<20.60$          &\\
%\hline
NGC\,5234     & 9.0  & ... & $<0.6$  & ... & $<0.07$ & $<1.2$ &  $<20.51$      & \\ 
%\hline
IC\,4312      & 18.7 & ... & $<0.8$  & ... & $<0.04$ & 4.2    &   21.14        & \\ 
%\hline
ESO\,221-20   & 9.9  & ... & $<0.3$  & ... & $<0.03$ &  $<0.93$      &     $<20.15$      & tail to
target from companion \\
             &       &     &    &     &      &        &           &  
CL: 4, 8, 16, ... $\times 10^{19}$ \\ 
\hline\hline
\end{tabular}
%\end{center}
\end{table*}
%\end{landscape} 

\eject
%\newpage

\begin{table*} 
\caption{Table2. Continued}
\begin{tabular}{lrcrccrrl} 
\hline\hline\\ 
Name &  $L_B$ & $S_{\HI}$&  $M_{\HI}$  & $\Delta V(20\%)$ &  $M_{\HI}$/$L_B$
&  $S_{\rm 1.4\ GHz}$ & log $P_{\rm 1.4\ GHz}$ & Morphology   \\ 
     & $10^9$ $L_\odot$ & Jy km s$^{-1}$ & $10^9$ $M_\odot$ & \kms &  &   mJy &
W Hz$^{-1}$   &Contour levels (CL) in   cm$^{-2}$ \\
\hline\\ 
ESO\,137-44A  &  ...    & ... & $<0.11$ & ... & ... &   15.2     &   21.80        &\\ 
%\hline
IC\,4647      &  ...    & ... & $<1.4$  & ... & ...   & $<1.6$     &   $<20.90$        &\\ 
%\hline
ESO\,103-49   & 6.2  & ... & $<0.8$  & ... & $<0.13$ & $<1.5$ &  $<20.82$         & \\  
%\hline
IC\,4847      & 5.3  & 4.03 &3.1     & 220 & 0.58 & $<1.0$ &  $<20.58$         & disk 
of $\sim 60$ kpc in size \\
             &       &     &         &     &      &        &           &
CL: 2, 4, 8, 16, ... $\times 10^{19}$ \\
%\hline
NGC\,6799     & 43.9 & 14.30 & 7.6    & 250 & 0.17 & $<1.0$ &  $<20.89$         & disk 
of
$\sim 240$ kpc in size, \\
             &       &       &     &     &      &        &           & possible 
90$^\circ$ warp \\
             &       &       &     &     &      &        &           & 
CL: 2, 4, 8, 16, ... $\times 10^{19}$ \\ 
%\hline
NGC\,6850     & 26.8 &  2.92 & 3.0    & 300 & 0.11 & $<1.5$ &  $<20.89$ & disk of
$\sim 100$ kpc\\
             &       &       &        &     &      &        &           & 
CL: 2, 4, 8, 16, ... $\times 10^{19}$ \\
%\hline
IC\,4970      & 13.4 &  ... & $<0.9$ & ... & $<0.07$ & $<1.2$ &  $<20.75$         & \\ 
%\hline
ESO\,234-11   & 13.8 &  ... & $<1.1$ & ... & $<0.08$ & $<1.0$ &  $<20.85$ & \\ 
%\hline
NGC\,6920     & 11.3 &  13.09 & 4.4    & 420 & 0.39 & 4.1    &  20.78    & disk of
$\sim 150$ kpc in size \\ 
             &       &        &        &     &      &        &           & 
CL: 2, 4, 8, 16, ... $\times 10^{19}$ \\
%\hline
NGC\,7166     & 13.5 & ...  & $<0.22$ & ... & $<0.02$ & $<1.4$ &  $<20.22$ &\\ 
%\hline
ESO\,240-10   & 27.0 & ...  & $<0.5$  &. .. & $<0.02$ & $<0.8$ &  $<20.22$ &\\
\\
\hline\hline
\end{tabular}
\\
References:
$^*$ from NVSS; 
$^1$ Oosterloo et al. (in prep); 
$^2$ Serra et al. (2006) 
\end{table*}

\vskip 2cm

\begin{table*} 
\caption{Early$-$type galaxies in our sample which have \HI\ imaging 
data  available in the literature and therefore were not re-observed. }
\setlength{\tabcolsep}{3pt} 
\begin{tabular}{llccrrcrccclcl}
\hline\hline\\ 
Name      &  &  RA        & Dec        &   $T$ & \multicolumn{1}{c}{v$_{\rm
hel}$}& $L_B$&   $M_{\HI}$ & $M_{\HI}$L/$_B$& $S_{1.4\ GHz}$ & log $P_{\rm
1.4\ GHz}$ &
 Morph. & Ref.  \\ 
          &  & \multicolumn{2}{c}{(J2000)}&   &  \kms &$10^9$\,$L_\odot$&
$10^9$\,$M_\odot$ &                & mJy    & W Hz$^{-1}$  &              &    \\
\hline \\
NGC\,802    &S0& 01 59 07.1 & $-$67 52 11  & $-$0.8&  1505 &     1.1         &
0.40             &   0.42      & {$<$} 1.9     &  \llap{$<$} 20.0  &disk  13 kpc   &  1 \\    
IC\,2006    &S0& 03 54 28.4 & $-$35 58 02  & $-$4.5&  1364 &     8.7         &
0.24             &   0.03      & {$<$} 3.0     & \llap{$<$} 20.1  &ring 30 kpc    &  2 \\
ESO\,118--34&S0& 04 40 17.2 & $-$58 44 47  & $-$2.0&  1171 &     1.1         &
0.13             &   0.12      &    4.0     &  20.1  &disk 7 kpc    &  1 \\ 
NGC\,1596   &S0& 04 27 37.8 & $-$55 01 37  & $-$2.0&  1465 &     8.5         &
0.8             &   0.03      & {$<$} 1.0     &  \llap{$<$} 19.7 &tail 40 kpc   &  3 \\  
NGC\,1947   &S0& 05 26 47.5 & $-$63 45 41  & $-$3.0&  1157 &     9.2         &
0.17             &   0.02      &   13.0     &  20.6  &disk  10 kpc   &  5 \\
NGC\,2328   &S0& 07 02 36.7 & $-$42 04 09  & $-$2.9&  1159 &     3.3         &
0.10             &   0.07      &    8.1     &  20.4  &disk  15 kpc   &  1 \\
NGC\,3108   &S0& 10 02 29.5 & $-$31 40 37  & $-$1.0&  2673 &    25.0         &
2.34             &   0.09      &  {$<$} 1.4    & \llap{$<$} 20.4  &disk   30  kpc  &  5 \\
NGC\,4936   &E & 13 04 17.1 & $-$30 31 34  & $-$5.0&  3215 &    35,3         &
1.1              &   0.03      &   39.8     &  22.0  &offset cloud 80 kpc &  9 \\  
NGC\,5266   &S0& 13 43 01.7 & $-$48 10 12  & $-$3.0&  2989 &    70.9         &
12.9              &   0.18      &   12.3     & 21.4   &disk 200 kpc    &  7 \\
NGC\,5291   &E & 13 48 49.1 & $-$30 13 07  & $-$4.6&  4376 &     6.3         &
1.6              &   0.25      &    5.8     &  21.4  &tail  180 kpc   &  8 \\
IC\,5063   &S0& 20 52 02.1 & $-$57 04 12  & $-$0.8&  3420 &    32.1         &
4.3              &   0.13      &   2100     &  23.8  &disk  30 kpc   &  5 \\
IC\,1459   & E& 22 57 09.5 & $-$36 27 37  & $-$5.0&  1836 &    34.0         &
0.2              &    0.06     &   1281     & 23.0  &tails\ $>200$ kpc &  6 \\
&      &      &         &     &            & \\
\hline\hline
\end{tabular}
\\
 References:
1) Sadler et al.\ 2000; 
2) Schweizer, van Gorkom \& Seitzer 1989; 
3) Chung et al.\ 2006 (astro$-$ph/0605600)
\\
4) Meurer, Staveley--Smith \& Killeen 1998; 
5) Oosterloo et al.\ 2002;
6) Oosterloo et al.\ 1999;
% unpublished ATCA data; 
%6a) Bureau et al., unpublished ATCA data; 
7) Morganti et al.\ 1997; \\
8) Malphrus et al.\ 1997; 
9) J.\ van Gorkom \& T.\ Oosterloo priv.\ comm.
%x) Morganti et al.\ 1998; 
%
%1) Oosterloo et al. 2002; 2) Oosterloo et al. 1999; 3) Koribalski
%et al. ?NGC\,3263??  4) Schiminovich et al.; 5) Morganti et al. 1997;
%2a) Franx, van Gorkom \& de Zeeuw 1994; 
\end{table*}

\end{document}